\documentclass{iopart}
\usepackage{amssymb}
\usepackage[dvips]{graphicx}

\begin{document}

\title[Network analysis for triggered searches]{Coherent network analysis for triggered gravitational wave burst searches }

\author{K Hayama$^1$, S D Mohanty$^1$, M Rakhmanov$^2$, S Desai$^2$}
\address{$^1$The University of Texas at Brownsville, 80 Fort Brown, 
Brownsville, Texas, 78520, USA.}
\address{$^2$ Department of Physics, Pennsylvania State University, University Park, PA 16802, USA.}
\ead{kazu@phys.utb.edu}

\begin{abstract}
Searches for gravitational wave bursts that are triggered by the observation of astronomical events 
require a different mode of analysis than all-sky, blind searches. For one, much more prior information is usually available in a triggered
search which can and should be used in the analysis. Second, since the data volume is usually small in a triggered search, it is 
also possible to use computationally more expensive algorithms for tasks such as data pre-processing that can
consume significant computing resources in a high data-volume un-triggered search. From the statistical point of view, the reduction in
the parameter space search volume leads to higher sensitivity than an un-triggered search. We describe here a data analysis pipeline for triggered
searches, called {\tt RIDGE}, and present preliminary results for simulated noise and signals. 
\end{abstract}

\section{Introduction}
Gravitational wave (GW) searches that are triggered by the observation of an astronomical event have some aspects that are fundamentally 
different from all-sky or un-triggered searches. First of all, the non-GW observations furnish a wealth of prior information that can guide the 
GW data analysis. For instance, in the case of a Gamma-ray Burst (GRB) trigger, we usually know the sky location and, sometimes, the distance to the
source. Based on the type of GRB, long-soft or short-hard, one can look for qualitatively different signals: short duration bursts for the former and binary inspiral signals for the latter. Second, there are many data pre-processing algorithms available in the literature that have been demonstrated to be useful
but have not been used in untriggered searches because of a natural selection effect: finite computational resources pick the computationally least expensive algorithm. Triggered searches, on the other hand, usually involve much smaller data volumes and, consequently, allow us the freedom to 
experiment with algorithms.

In this paper, we describe results from one such ongoing experiment of using new algorithms for analyzing data from the LIGO, GEO and Virgo detectors
for triggered searches. We use new algorithms for both the data pre-processing ({\em conditioning}) and the search part of the analysis pipeline. The code
for the pipeline, called {\tt RIDGE}, is written mostly in matlab. We will not describe the code in any detail in this paper but focus on the algorithms and
results.

\section{Data analysis components}
A typical triggered search involves several components~\cite{sdm_amaldi2002}. First, a mechanism is required to collect trigger information. This is already well developed and nearly automated within the LIGO Scientific Collaboration (LSC), at least as far as GCN~\cite{GCN} triggers are concerned. Second, given the wealth of astrophysical prior information associated with most triggers, we need proper mathematical 
tools to incorporate this information into our analysis. So far, this aspect of a triggered search has seen only limited advances. 
The most detailed use of prior information has been in the search~\cite{sgr1806search} for GWs from the SGR-1806--20 event~\cite{sgr1806event} where X-ray intensity oscillations associated with the flare were used to motivate a 
search for quasi-monochromatic signals at specific frequencies. 
However, more work is needed in this area, especially on the use of Bayesian approaches modeled 
after  the ones used for a different type of ``triggered" search~\cite{known_pulsar_searches}, namely GWs from known pulsars in the Galaxy. The other components required in a typical triggered search, as in any other search,
 are {\em data conditioning} and the generation of a detection statistic. 

%%%%%%%%%%%%%%%%%%%%%%%%
\subsection{Data conditioning in {\tt RIDGE}}
\begin{figure}
\begin{center}
\includegraphics[scale = 0.5]{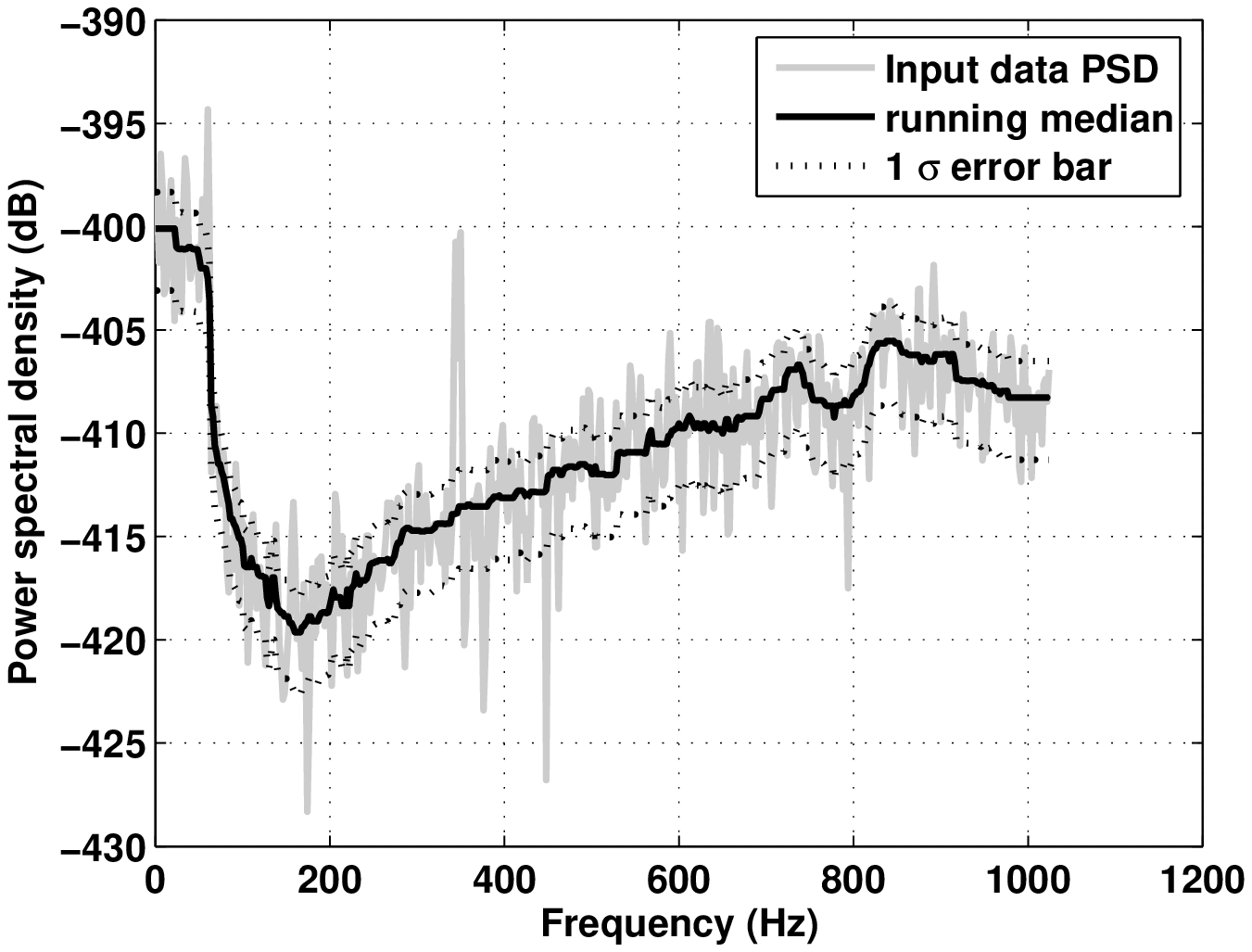}
\includegraphics[scale = 0.5]{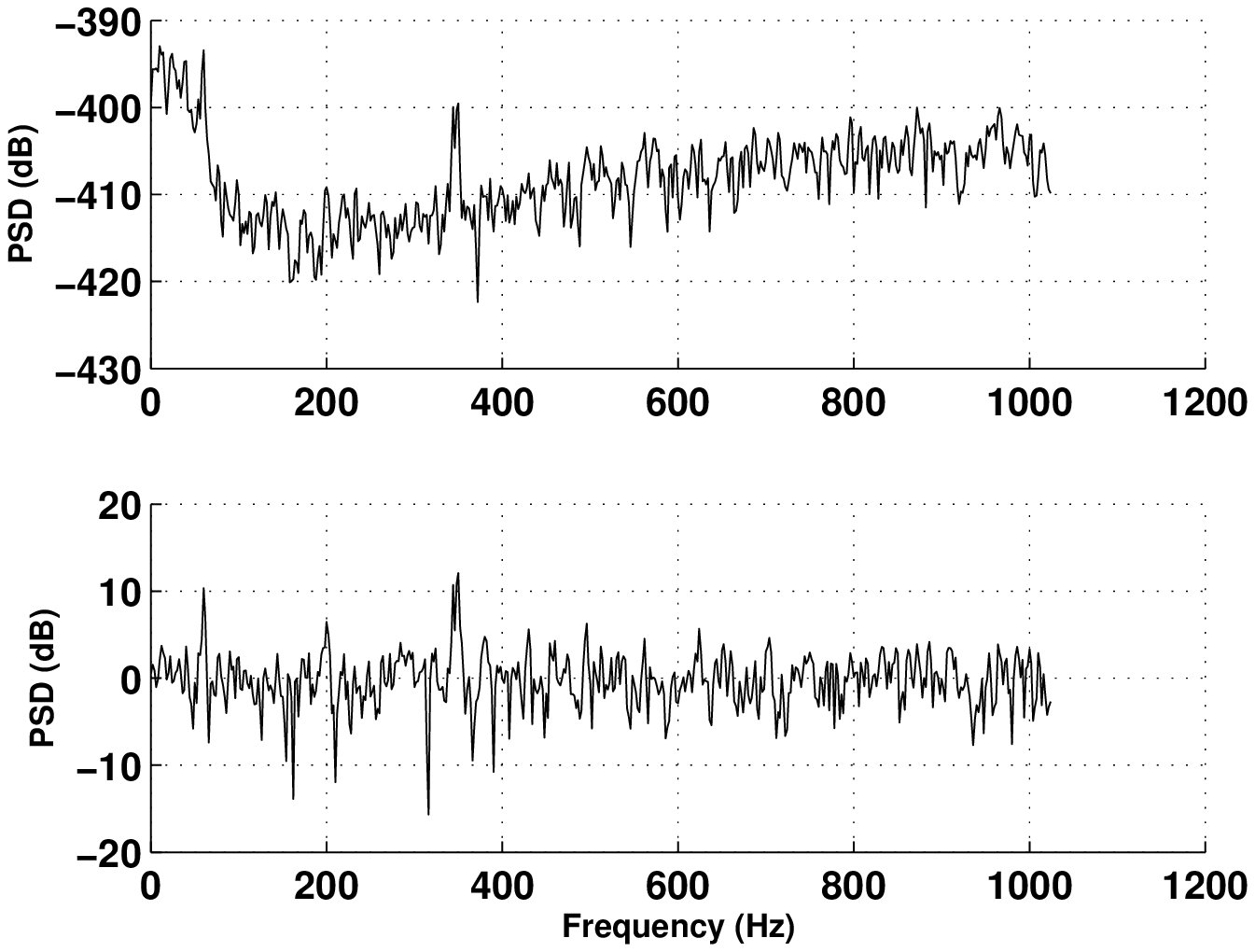}
%\end{center}
\caption{
A sample power spectral density and the corresponding running median estimate of the noise (top plot). Note that the noise floor estimation was not 
affected by the large outlier due to a strong sinusoidal signal. The plot at the bottom shows the psd of raw data (top) and the psd (bottom) of the data after passing it through the whitening filter. The whitened data has variance 1, which explains the change in level to 0 dB in the bottom plot.
\label{fig_runningmedian}
}
\end{center}
\end{figure}

The purpose of data conditioning is to mitigate the effect of known instrumental artifacts in the data. In {\tt RIDGE} this is done in a two step process. First,
the {\em noise floor} of the power spectral density (PSD) of noise is estimated using a running median~\cite{sdm:MBLT}. Fig.~\ref{fig_runningmedian} shows the PSD of 2~sec of simulated detector noise with a power spectral density approximating the LIGO SRD curve for the 4km detectors. 
Superimposed on the estimated PSD is the running median estimate of the noise floor. The main advantage of the running median is its ability to reject outliers. In this particular case, the outliers are the narrowband noise artifacts arising from sources such as power line interference (modeled as 
sinusoids in this particular example).

 If $S(f)$ denotes the running median estimate of the PSD, a digital filter with a transfer function $T(f)$ such that
$|T(f)| = 1/\sqrt{S(f)}$ will {\em whiten} the data. Fig.~\ref{fig_runningmedian} also shows the PSD of the output of such a digital filter.
In {\tt RIDGE}, the whitening filters are Finite Impulse Response (FIR) filters and the
filter coefficients are obtained using the PSD obtained 
from a user-specified training data segment. The digital filter is then applied to a longer stretch of data. It is clear 
 that any non-stationarity in the noise floor on timescales shorter than the gap between the training and the processed data
 can create a discrepancy in the trained filter and the filter that is required 
for whitening the longer stretch. 
To the knowledge of the authors, not much concrete work exists in the GW literature on handling this issue robustly in any analysis pipeline. However, tools already exist that can monitor the behaviour of the noise floor in real time~\cite{noisefloormon}, and characterize burst-like non-stationary components~\cite{monnonstat}.  We intend to integrate these 
information into our pipeline in the future.

\begin{figure}
\includegraphics[scale=0.5]{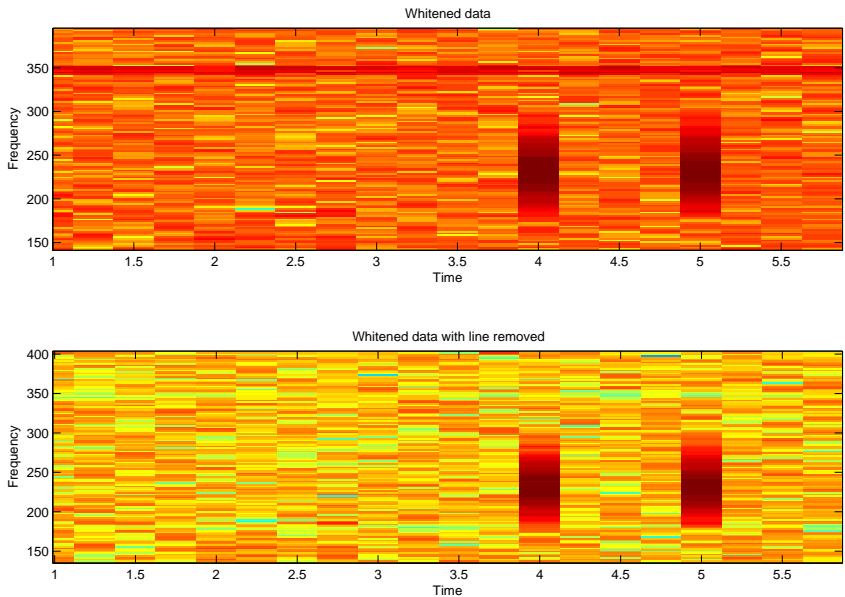}
\caption{
Spectrograms of simulated data with (top) line present in the data along with two transient signals and (bottom) same data with the line 
removed using {\tt MBLT}. This simulation actually consisted of more lines but due to the limitations of the graphics used, we show only a 
zoomed in view of a particular region of the time-frequency plane. The simulated line feature appears as the straight horizontal line in the upper plot
near 350~Hz; the feature actually consists of two closely spaced sinusoids at 344 and 349~Hz.
\label{specgrams}
}
\end{figure}

The whitening step above is followed by a line estimation and removal step which reduces or, in some cases, eliminates the effect of 
strong narrowband noise features ({\em line} noise) arising from power line interference or violin modes. The method used here, called {\tt MBLT},
 is described 
in~\cite{sdm:MBLT}. Essentially, the method consists of estimating the amplitude and phase modulation of a line feature at a given 
carrier frequency (which have to be found manually). Fig.~\ref{specgrams}
 shows spectrograms of the whitened data before and after line removal. In this example, signals were injected into the data
prior to line removal. As can be seen, the line removal does not affect the signals in any significant way. This is an inbuilt feature of {\tt MBLT} which uses the 
running median for estimation of the line amplitude and phase functions. Transients signals appear as outliers in the amplitude/phase time series
and are rejected by the running median estimate. The whitened and line removed data is then passed on to the next step of {\tt RIDGE}.

 %%%%%%%%%%%%%%%%%%%%%%%%
\subsection{Detection statistic}

The basic algorithm implemented for coherent network analysis in {\tt RIDGE} is {\em regularized}
 maximum likelihood, described in
several recent papers~\cite{klimenko+etal,rakhmanov,mohanty+etal}. The GW response of any one detector is a linear combination of the two 
unknown polarization waveforms $h_+(t)$ and $h_\times(t)$ arriving at the detector from a certain direction $\theta_0$, $\phi_0$ on the sky. We assume the  Earth-centered, ecliptic reference frame here for defining the polar angle, $\theta$, and azimuthal angle, $\phi$. It was shown in~\cite{klimenko+etal}
that the problem of inverting a set of detector responses to obtain $h_+(t)$ and $h_\times(t)$ is an {\em ill-posed} one and needs to be solved by
using some kind of regularization. Alternatives to the regularization used in~\cite{klimenko+etal} were proposed in~\cite{rakhmanov,mohanty+etal}.
The theoretical aspects of regularization related to the direction dependent 
antenna patterns of the detectors in a network is fairly well understood now.

The particular regularization scheme used to obtain the results reported here was developed in~\cite{rakhmanov}. 
It is quite easily shown that the {\em hard constraint}
algorithm presented in~\cite{klimenko+etal} is actually obtained in the limit of an infinitely large gain factor for the regulator
of~\cite{rakhmanov}. The input to the algorithm is a set of equal length,
conditioned data segments from the detectors in a given network. The output, for a given
sky location $\theta$ and $\phi$, is the
value of the likelihood of the data maximized over all possible $h_+$ and $h_\times$ waveforms with durations less than or equal to the data segments.
The maximum  likelihood values are obtained as a function of $\theta$ and $\phi$ -- this two dimensional output, ${\bf S}(\theta,\phi)$,
 is called a {\em sky-map}. 

\begin{figure}
\begin{center}
\includegraphics[scale=0.7]{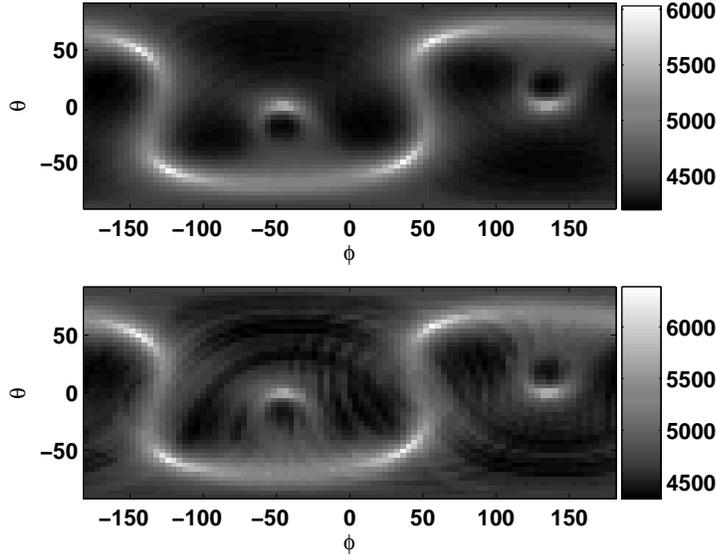}
\caption{
Plots of the sky-map for noise only case (top) and signal plus noise case (bottom). The detector network used was H1, H2, L1 and V1.
The top plot was made by averaging the sky-maps for 100 noise 
realizations. The bottom plot  is an average over sky-maps made with 4 independent realizations of noise plus a signal with
$h_{\rm rss} = 5\times 10^{-22}$~${\rm Hz}^{-1/2}$ (c.f., Eq.~\ref{hrss_def}). 
The magnitude scale is
in arbitrary units. The test statistics are constructed by normalizing the maps.
\label{skymaps}
}
\end{center}
\end{figure}

Fig.~\ref{skymaps} shows examples of sky-maps obtained for the network consisting of the two 4km LIGO interferometers and the 2km LIGO interferometer. 
As can be seen from the figure, the presence of a signal can ``ring off" the entire sky. As the direction to a trigger will usually be
well known in advance from electromagnetic observations, say, there is no necessity in a triggered search 
to restrict ourselves only to a part of the sky-map for the purpose of detection. We can construct a detection statistic using the 
entire sky map. In the present paper, we consider two different choices. The obvious one is (1),
\begin{equation}
R_{mm} = \frac{\max_{\theta,\phi} {\bf S}(\theta,\phi)}{\min_{\theta,\phi} {\bf S}(\theta,\phi)}\;.
\end{equation}
$[\max/\min]_{\theta,\phi}{\bf S}(\theta,\phi)$ represents a [maximum/minimum] value of ${\bf S}(\theta,\phi)$ on the $\theta$--$\phi$ plane. A less obvious choice is (2),
\begin{eqnarray}
R_{\rm rad} &=& \left[  \left(\frac{\max_{\theta,\phi} {\bf S}(\theta,\phi)}{\max_{\theta,\phi} \overline{\bf S}_0(\theta,\phi)}-1\right)^2 +\right. \nonumber\\
&&\left. \left(R_{mm}\frac{\min_{\theta,\phi} \overline{\bf S}_0(\theta,\phi)}{\max_{\theta,\phi} \overline{\bf S}_0(\theta,\phi)}-1\right)^2 \right]^{1/2} \;,\\
\overline{\bf S}_0(\theta,\phi) & = & {\rm E}\left[ {\bf S}(\theta,\phi) | \mbox{no signal in data}\right] \;,
\end{eqnarray}
where ${\rm E}[{\bf S}(\theta,\phi)|  \mbox{no signal in data}]$ denotes an ensemble average taken over realizations of sky-maps that do not have any GW signals.
The second statistic, $R_{\rm rad}$, is the radial distance of the observed values, in an 
appropriately scaled $( R_{mm}$, $\max_{\theta,\phi} {\bf S}(\theta,\phi) )$ plane,
from the mean location of the same quantities in the absence of a signal.
%%%%%%%%%%%%%%%%%%%%
\section{Results}
\begin{figure}
\begin{center}
\includegraphics[scale=0.5]{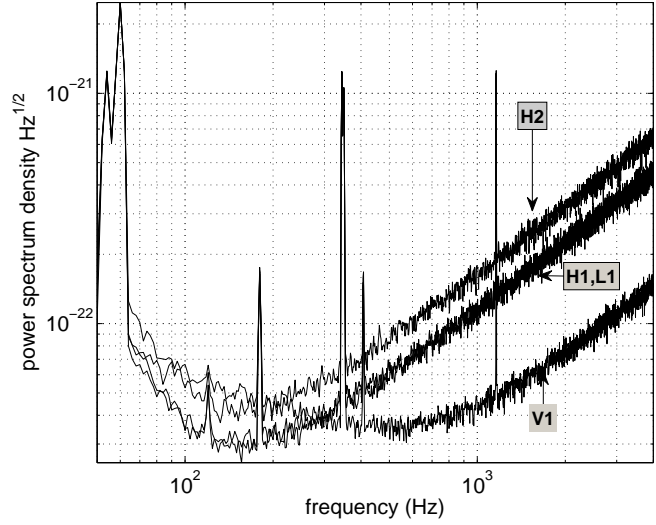}
\caption{Power Spectral Densities of simulated detector noise for H1 (L1), H2 and V1. The power spectral densities were computed with a frequency
resolution of $2.0$~Hz.
\label{simpsds}}
\end{center}
\end{figure}

We carried out Monte Carlo simulations to characterize the performance of the detection statistics $R_{mm}$ and $R_{\rm rad}$. The network consisted
of the 4km LIGO detectors (H1 and L1), the 2km LIGO detector (H2) and the Virgo detector (V1). For the detector 
noise PSD, we used the design sensitivity 
curves for the LIGO and Virgo detectors as given in~~\cite{LIGOdesignsens,Virgodesignsens}
 and kept the locations and orientations the same as the 
real detectors. Gaussian, stationary noise was generated ($\sim 3000$~sec) by 
first generating 4 independent realizations of white noise and then passing them through FIR filters having
transfer functions that {\em approximately} match the design curves.
To simulate instrumental artifacts, we added sinusoids with large amplitudes at $(54, 60, 120, 180, 344, 349, 407)$~Hz. 
The resulting PSDs of the simulated data are shown in Fig.~\ref{simpsds}. 

\begin{figure}
\begin{center}
\includegraphics[scale=0.6]{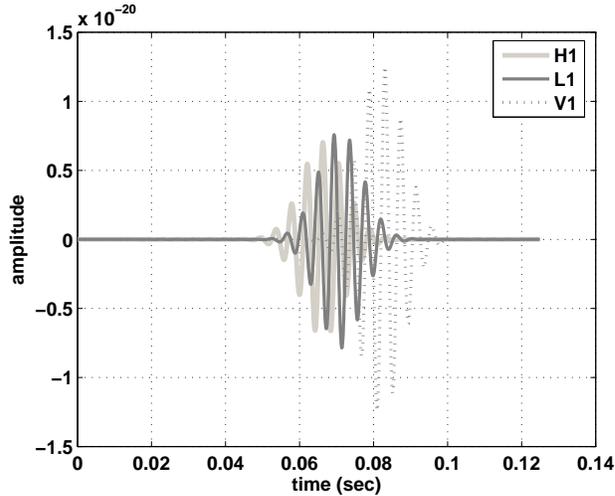}
\caption{The detector responses to the injection signal used in the Monte Carlo simulations. The strain response for H1 and H2 are identical because
of their co-location and co-alignment.
\label{detresponses}
}
\end{center}
\end{figure}

Signals of constant amplitude were added to the simulated noise at regular intervals.
The injected signals corresponded to a single source located at
$\theta = -30^\circ$ and $\phi = 167.0^\circ$, which was the location of a real trigger, GRB060813. We assumed that 
the $h_+(t)$
and $h_\times (t)$ waveforms were both sine-Gaussian signals having center frequency $235$~Hz  and a $Q = 9$. The $h_\times(t)$ waveform
has an added phase of $\pi/2$ relative to the phase of $h_+(t)$. Fig.~\ref{detresponses}
shows the signals, i.e., the detector responses, that were added to the noise. 
The signal strength is specified in terms of the $h_{\rm rss}$ value, defined as
\begin{equation}
h_{\rm rss} = \left[\int_{-\infty}^{\infty} dt\; \left(h_+^2(t) + h_\times^2(t)\right) \right]^{1/2}\;.
\label{hrss_def}
\end{equation}
In our results we use $h_{\rm rss}$ values of $3\times 10^{-22}$, $5\times 10^{-22}$ and $7\times 10^{-22}$~${\rm Hz}^{-1/2}$.

The simulated data were generated at a sampling frequency of 16384~Hz and then passed through the data conditioning
pipeline. Besides downsampling the data by applying the same anti-aliasing filter to all data streams, 
the data conditioning pipeline applies time domain whitening 
filters that were trained, as described earlier, on the first two seconds of data for
each detector (without any injected signals). The conditioned data is then passed to the sky-map generation code in segments of length 0.5~sec.
This integration time is much larger than the $\sim 40$~msec duration of the injected signals and, hence, the performance estimates we obtain 
here should be taken as lower bounds. The reason behind the large choice of 0.5~sec was to accomodate the filter delays introduced into the data.
In a future version of the pipeline, these delays, which are known since the filters are all FIR filters, will be removed carefully. This will allow us to 
shrink the integration time for short duration signals.

\begin{figure}
\begin{center}
\includegraphics[scale=0.5]{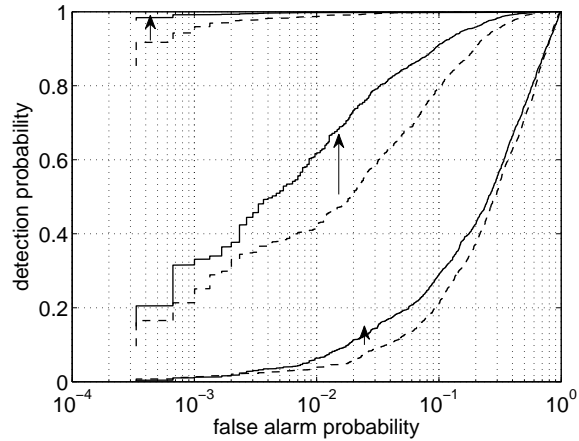}
\caption{Receiver Operating Characteristics for $R_{mm}$ (dashed) and $R_{\rm rad}$ (solid) for $h_{\rm rss} = 3\times 10^{-22}$ (bottom), 
$5\times 10^{-22}$ (middle) and $7\times 10^{-22}$~${\rm Hz}^{-1/2}$ (top). The enhancement when going from $R_{mm}$ to $R_{\rm rad}$
is shown by the arrows. 
\label{roc_reggain_0p2}}
\end{center}
\end{figure}

Fig.~\ref{roc_reggain_0p2} shows the Receiver Operating Characteristic (ROC) curve for the simulation described above. ROC curves for both $R_{mm}$
and $R_{\rm rad}$ are shown. Clearly, the $R_{\rm rad}$ statistics performs better than the more naive choice $R_{mm}$, demonstrating
that a judicious choice of a statistic that incorporates information from the full sky-map is definitely required for triggered searches. 
We believe that more 
effective full sky-map statistics can be derived that can push performance to even better levels. For the current simulation and choice of parameters,
we find that an $h_{\rm rss}$ of 
$5 \times 10^{-22}$~${\rm Hz}^{-1/2}$ can be detected with $\simeq 62\%$ probability at a false alarm probability of $10^{-2}$. The false 
alarm probability corresponds to a rate of 1 false event in $50$~sec.
%%%%%%%%%%%%%%%%%%%%
\section{Conclusion}
We describe here an analysis pipeline, {\tt RIDGE}, that implements regularized maximum likelihood based network analysis for triggered searches.
Monte Carlo simulations with simulated detector noise show that the pipeline is quite sensitive: For a LIGO-Virgo network,
a $\simeq 62\%$ detection probability is obtained 
for a signal $h_{\rm rss} = 5\times 10^{-22}$~${\rm Hz}^{-1/2}$ at a false alarm rate of $2\times10^{-2}$~Hz. 
All the algorithms used in this pipeline are significantly different from other network analysis 
pipelines~\cite{Cwb,Xpipeline} that are also currently being applied to
 triggered searches. (The basic statistic used in~\cite{Xpipeline} is the 
hard constraint statistic introduced in~\cite{klimenko+etal}.) Thus, at both the level of algorithms and the associated
software, {\tt RIDGE} should provide an independent check for results obtained with other pipelines. {\tt RIDGE} is also being designed to
serve as a platform for trying out new regulators. In future work, since {\tt RIDGE} pipeline does not only extract candidates of gravitational wave signals, but also reconstruct  $h_+$ and $h_{\times}$ waveform at arbitrary sky location, we integrate a waveform estimation method~\cite{west} and investigate how accurately astrophysical information can be obtained.

%%%%%%%%%%%%%%%%%%%%
\ack{K.H. is supported by NASA grant NAG5-13396 to the Center for  Gravitational Wave Astronomy at the University of Texas at Brownsville.  SDM's work was supported by NSF grant PHY-0555842.  The work of S.D. and M.R. were supported by
 the Center for Gravitational Wave Physics at the Pennsylvania State
 University and the US National Science Foundation under grants
 PHY 00-99559, PHY 02-44902, PHY 03-26281, and PHY 06-00953. The Center for Gravitational Wave
 Physics is funded by the National Science Foundation under
 cooperative agreement PHY 01-14375.  This paper has been assigned LIGO Document Number LIGO-P070045-01.
}
%%%%%%%%%%%%%%%%%%%%%


\section*{References}
\begin{thebibliography}{99}
\bibitem{sdm_amaldi2002}
S D Mohanty, S Marka, R Rahkola, S Mukherjee, I Leonor, R Frey, J Cannizzo and J Camp,
\ 2004, Class. Quantum Grav. 21 No 5  S765-S774
\bibitem{GCN} Global Coordinates Network, http://gcn.gsfc.nasa.gov
\bibitem{sgr1806search}
The LIGO Scientific Collaboration, \ 2007, LIGO-P040055-01-Z
\bibitem{sgr1806event}
Mereghetti, S., et al., \ 2005, A \& A, 433-2, L9-L12 
\bibitem{known_pulsar_searches}
R. J. Dupuis, G. Woan,
\ 2005, Phys.Rev. D72 102002
\bibitem{sdm:MBLT}
S D Mohanty, \ 2002 Class. Quantum Grav. 19 No 7 1513-1519
\bibitem{noisefloormon}
S Mukherjee, \ 2003 Class. Quantum Grav. 20 No 17 S925-S936
\bibitem{monnonstat}
K. Hayama and M.-K. Fujimoto
\ 2006 Class. Quantum Grav. 23 S9-S15
\bibitem{klimenko+etal}
S. Klimenko, S. Mohanty, M. Rakhmanov, G. Mitselmakher
\ 2005 Phys.Rev. D72 122002
\bibitem{rakhmanov}
M. Rakhmanov \ 2006 Class.Quant.Grav. 23 S673-S686
\bibitem{mohanty+etal}
S. D. Mohanty, M. Rakhmanov, S. Klimenko, G. Mitselmakher
\ 2006 Class.Quant.Grav. 23 4799-4810
\bibitem{LIGOdesignsens}
\begin{verbatim}http://www.ligo.caltech.edu/~jzweizig/distribution/LSC_Data\end{verbatim}
\bibitem{Virgodesignsens} 
\begin{verbatim}http://www.virgo.infn.it\end{verbatim}
\bibitem{Cwb}
S. Klimenko
\ 2006 G060565-00-Z
\bibitem{Xpipeline}
S. Chatterji, A. Lazzarini, L. Stein, and P. J. Sutton, A. Searle, and M. Tinto,
\ 2006, Phys. Rev. D 74, 082005, 
P J. Sutton and M Was,
\ 2007 LIGO-T070091-00-Z
\bibitem{west}
K. Hayama
\ 2004 Progress of Theoretical Physics 111 vol.6 807
\end{thebibliography}
\end{document}